\documentclass[5p]{elsarticle}
\usepackage{epstopdf}
\usepackage{lineno,hyperref}
\modulolinenumbers[5]
\journal{Journal of the European Ceramic Society}

\begin{document}
\begin{frontmatter}
\title{Structural, magnetic, dielectric and mechanical  properties of (Ba,Sr)MnO$_3$ ceramics}

\author[UP]{R.~Bujakiewicz-Koronska\corref{cor1}}
%\fnref{fn2}}
\ead{sfbujaki@cyf-kr.edu.pl}
\author[UP]{D.~M.~Nalecz} %\fnref{fn2}}
\ead{sfnalecz@cyf-kr.edu.pl}
\author[IFUJ]{A.~M. Majcher} %\fnref{fn2}}
\ead{anna.majcher@uj.edu.pl}
\author[IFJ]{E.~Juszynska-Galazka}
\ead{Ewa.Juszynska-Galazka@ifj.edu.pl}
\author[IFJ]{M.~Galazka} %\fnref{fn2}}
\ead{Miroslaw.Galazka@ifj.edu.pl}
\author[LP]{L.~Vasylechko} %\fnref{fn2}}
\ead{crystal-lov@polynet.lviv.ua}
\author[IFM]{E.~Markiewicz} %\fnref{fn2}}
\ead{ewamar@ifmpan.poznan.pl}
\author[WCUJ]{D.~Majda} %\fnref{fn2}}
\ead{dorota.majda@uj.edu.pl}
\author[LU]{A.~Kalvane} %\fnref{fn2}}
\ead{kalvane@cfi.lu.lv}
\author[IF]{K.~Koronski} %\fnref{fn2}}
\ead{koronski@ifpan.edu.pl}

\address[UP]{Institute of Physics, Pedagogical University, Podchorazych 2, PL-30-084 Cracow, Poland}
\address[IFUJ]{Institute of Physics, Jagiellonian University, Prof. Stanislawa Lojasiewicza 11, PL-30-348 Cracow, Poland }
\address[IFJ]{Niewodniczanski Institute of Nuclear Physics PAN ul.Radzikowskiego 152, PL-31-342 Cracow, Poland }
\address[LP]{Lviv Polytechnic National University, 12 Bandera St., 79013 Lviv, Ukraine }
\address[IFM]{Institute of Molecular Physics, Polish Academy of Science, Smoluchowskiego 17, PL-60-179 Poznan, Poland }
\address[WCUJ]{Department of Chemistry, Jagiellonian University, Ingardena 3, PL-30-060 Cracow, Poland }
\address[LU]{Institute of Solid State Physics, University of Latvia, 8 Kengaraga str., LV-1063, Riga, Latvia}
\address[IF]{Institute of Physics PAS, Al. Lotnikow 32/46, PL-02-668 Warsaw, Poland}
\cortext[cor1]{Corresponding author}

\begin{abstract}
Ceramic samples, produced by conventional sintering method in ambient air, 6H-SrMnO$_3$(6H-SMO), 15R-BaMnO$_3$(15R-BMO),  4H-Ba$_{0.5}$Sr$_{0.5}$MnO$_3$(4H-BSMO)  were studied. In the XRD measurements  for SMO the new anomalies of the lattice parameters at 600-800 K range and the increasing of thermal expansion coefficients with a clear maximum in a vicinity at 670 K were detected. The  N$\acute{e}$el phase transition for BSMO was observed at $T_N$=250 K in magnetic measurements and its trace was detected in dielectric, FTIR, DSC, and DMA experiments. The enthalpy and entropy changes of the phase transition for BSMO at $T_N$ were determined as 17.5 J/mol and 70 mJ/K mol, respectively. The activation energy values and relaxation times  characteristic for relaxation processes were determined from the Arrhenius law.  Results of {\it ab initio} simulations showed that the contribution of the exchange correlation energy to the total energy is about 30\%. 
\end{abstract}

\end{frontmatter}

%\linenumbers
%%%%%%%%%%%%%%%%%%%%%%%%%%%%%%%%%%%%%%

\section{Introduction}
\label{sec1}
Since in 1994 Schmid \cite{1} introduced the concept of so-called multiferroics systems into scientific space huge progress is observed in discovering new
 properties and applications possibilities of new intelligent materials \cite{2}.  The mechanisms governing multiple ferroic orders in these materials and the coupling between them belong to the frontier research  in physics  \cite{3,4,5,6,7,8,9,10}. Couplings between magnetic, electric and elastic properties in multiferroics seem immediate in the expansion  of a free energy in the function of fields up to second or third order \cite{6,8}. Derivatives of the free energy with respective field give macroscopic order parameters \cite{9} which can be used to describe these couplings. Their experimental values allow for classiffication of multiferroic materials into I$^{st}$, II$^{nd}$ or III$^{rd}$ group \cite{7,8}. Out of many types of multiferroics the manganites have gathered much interest because of their marked magnetoelectric response. The group of hexagonal manganites $R$MnO$_3$, where $R$  is  rare earth (Er, Ho,  Lu, Tm, Yb, Y, Sc, \cite{10,11}), indium or alkaline elements (Ba, Sr, \cite{12,13,14,15, Neg-70, Neg-73, Dab-03, Pra-14, Goi-16, Trok-16}), show fascinating properties due to their promising  wide applications. 
 
The different hexagonal polymorphs of (Ba,Sr)MnO$_3$ have  not been proved to show any ferroelectric or polar order. There are some
theoretical reports \cite{Var-13} which predict that hexagonal 2H-BaMnO$_3$ might be ferroelectric. However, the
ferroelectric mechanism in 2H-BaMnO$_3$ is completely different from that of perovskite (Sr,Ba)MnO$_3$.
The ferroelectric order in the perovskite phase is similar to prototype ferroelectric BaTiO$_3$, i.e. off-centring of B cation (Ti or Mn). In 2H-BaMnO$_3$ the ferroelectric mechanism is improper and similar to that one reported in hexagonal YMnO$_3$ \cite{Aken}. Thus, the ferroelectric properties (polarization values, dielectric permittivity, Curie temperatures, etc.) differ enormously.

The stability of the perovskite phase in (Sr,Ba)MnO$_3$, especially in Ba-rich solid solutions, is extremely difficult  \cite{Dab-03} or \cite{Neg-73} and only a few research groups have managed to synthesize it \cite{13,Lan-15}.
%(see for instance Ref. 13 for bulk specimens or Langenberg et al. ACS Appl. Mater. Interfaces 7, 23967 (2015) for thin films). 
The different hexagonal polymorphs are the ground state and are very stable \cite{Niel-14}. 
Alkaline manganites (Ba,Sr)MnO$_3$ crystals show strong spin-phonon coupling \cite{14}, ferroelectric Curie temperature about 410~K, N$\acute{e}$el temperature about 217-230~K \cite{14,15}.

In this article we have focused on the Ba$_x$Sr$_{1-x}$MnO$_3$ ceramic system. Recently, impedance spectroscopy measurements for Ba$_x$Sr$_{1-x}$MnO$_3$, $x = 1.0, 0.5$ and $0.0$ have been published \cite{16}. Alkaline manganites ceramics BaMnO$_3$ (BMO), Ba$_{0.5}$Sr$_{0.5}$MnO$_3$ (BSMO), and SrMnO$_3$ (SMO) synthesized by conventional high-temperature method in ambient air were studied by several experimental methods in a wide temperature range.

The details of our measurements are given in Sec. \ref{sec2}. The experimental results are presented and discussed in Sec. \ref{sec3}. Results of theoretical \textit{ab initio} simulations in Sec. \ref{sec4}  are shown. Finally, the summary is presented in Sec. \ref{sec5}.

\section{Experimental}
\label{sec2}

Ceramic samples were produced by conventional method in ambient air with heating/cooling rate at 300 K/h  and sintering temperature T$_{sint}$ as follows: BMO at T$_{sint}$=1293~K per 5h, firing at 1603~K per~4h,
BSMO at T$_{sint}$=1323~K per 5h and sintered at 1313 K per~3h,
SMO at T$_{sint}$=1273~K per 5h and sintered at 1543~K per~4h (details are described in \cite{16}). The Sigma Aldrich components BaCO$_3$, SrCO$_3$, MnO$_2$ with purity 99.0 \% were used.

  BMO and SMO ceramics received densities on the level 72 \% of the theoretical ones. BSMO achieved 86 \% of the theoretical density.

	For phase and structural characterization of the samples at room temperature (RT) X-ray powder diffraction measurements were performed on the laboratory Huber imaging plate Guinier camera G670 (CuK$_{\alpha 1}$ radiation,  $\lambda$ =1.54056 \AA). The diffraction patterns were registered in the 2$\theta$ range 5$^\circ$ - 100$^\circ$ with  a scan step of  0.005$^\circ$. The lattice parameters, fractional coordinates of the atoms and their displacement parameters were refined by full-profile Rietveld method implemented \  in the \  program package WinCSD~\cite{17}. Structural investigation and thermal behaviour of the crystal structure of SMO in the temperature range of 298-1173 K was carried out by means of high-resolution X-ray powder diffraction technique applying synchrotron radiation. \textit{In situ}  high-temperature diffraction experiments were made at the beamile ID22 of ESRF (Grenoble, France) during the beam time allocated to the ESRF Experiment N$^\circ$~MA-2320.

Microstructures of the  samples were observed using scanning electron microscope (SEM) Hitachi SU70 equipped with an X-ray micro-analysis system. The samples surfaces additionally were polished and etched by 95\% H$_2$O+ 4\% HCl+1\% HF for 4 min.  

Magnetic measurements were performed for the BMO, BSMO, SMO samples using the Quantum Design MPMS 5XL SQUID magnetometer in a wide range of temperatures and field (2--300 K, 0--50 kOe). The temperature dependences of the static susceptibility $\chi_{DC}$ in ZFC (zero-field-cooled) and FC (field-cooled) mode were measured in the external magnetic field H equal to 50~Oe.

	Differential scanning calorimetry (DSC) measurements were carried out using a DSC 821e Mettler Toledo apparatus with an intra-cooler unit (Haake) in heating process from 120 K to RT with cooling/heating rate 2~K/min. All samples were closed in an aluminum pan hermetically sealed with an aluminium lid after annealing at 600 K by 3 h. The DSC cell was rinsed out by argon (80 cm$^{3}$/min) during the experiment. The masses of the samples were about 30 mg. The DSC measurements for all samples were taken two times because of the fact that  the temperature history of the sample might affect the result. The second curve was treated as the final outcome.

	The Fourier transform infrared absorption (FT-IR)  spectra in the wave number range of 400 -4000~cm$^{-1}$ were carried out using an EXCALIBUR FTS-3000 spectrometer in the temperature range 20--300~K and measured with the sample mixed with KBr. The sample was  sandwiched between two KRS-5 window disks. During the experiments, the spectrometer was purged with dry nitrogen. Absorption measurements were conducted during cooling and heating.
 
	Dielectric properties of the BMO, BSMO and SMO ceramics were studied using the equipment described in \cite{16} on heating process at a rate of 1 K/min  in the temperature range 125--300~K and frequencies 1--400 kHz. The impedance spectra of the samples were taken within the same temperature range and at the frequency varied from 1 Hz to 1 MHz using the same equipment. The data were collected and estimated by a WinDATA impedance analysis software and WinFit V 3.2. program.

	Dynamical mechanical analysis (DMA) was carried out using a PerkinElmer DMA 8000 instrument. Solid steel material pockets were obtained from PerkinElmer (Seer Green, U.K., Part No. N5330323). BMO, BSMO and SMO samples were grinded and then plated in the steel pockets. The masses of the samples placed into the pockets were about 50 mg. The grains sizes were of the order of several micrometers. The measured values of $E'$ were storage moduli of the steel pockets and grinded samples. The contribution from the grinded samples depended on their Young modulus. The observed changes in slopes of $E'$ as a function of the temperature detect changes in hardness and stiffness of the material inside the pocket. Therefore, this kind of measurements can be treated as qualitative type.The pockets were folded in half, crimped closed to form a sandwich and clamped into the DMA. The pocket was located in a single cantilever bending mode, with one end of the pocket clamped to a fixed support and the other end clamped to the drive shaft (subjected to an oscillating displacement by the drive shaft) \cite{18,19}. This resulted in the pocket being deformed in an oscillating, bending motion in and out of plane, subjecting the sample powder in the pocket to a horizontal shear. The samples were cooled down to about 173~K and heated with a rate of 3 K/min, while undergoing a dynamical displacement of about 0.05 mm at 1 Hz. The distances between the fix and oscillating clamps were 15.04, 14.78 and 15 mm for BMO, BSMO, and SMO, respectively. The widths of  the folded steel pockets with the samples were 7.62, 7.60 and 7.56 mm for BMO, BSMO, and SMO, respectively. The thicknesses of the pockets with the samples were 0.75 mm (BMO), 0.74 mm (BSMO), and 0.72 mm (SMO). The thicknesses of empty folded pockets were 0.5 mm. The errors of measured free lengths, widths and thicknesses are about 0.02 mm. The measured storage modulus $E'$ for dynamical displacement equalling about 5\% of the thickness of the sample can be treated as 'effective' Young modulus of the composite in the steel pocket.

	For a complementary description of electronic properties of the studied materials we performed \textit{ab initio} simulations. The Spanish Initiative for Electronic Simulations with Thousands of Atoms (SIESTA) method was used for calculations of electronic structure for BMO, BSMO, and SMO materials \cite{20,21}. The calculations of the density functional theory (DFT) were carried out using the generalized-gradient approximation (GGA), the Perdew– Burke-Ernzerhof (PBE) parametrization and the spin polarization inside the SIESTA 3.2 package \cite{22,23,24}. 

\section{Results and discussion}
\label{sec3}
\subsection{XRD measurements}

\begin{figure}
\centering
\includegraphics[width=0.4\textwidth]{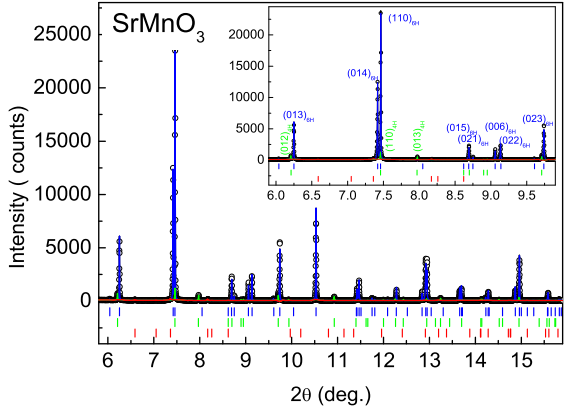}
\caption{Results of multi-phase Rietveld refinement of X-ray synchrotron powder diffraction pattern ($\lambda$ = 0.35434 \AA) showing the presence of  89.1 wt.\% of 6H-SMO (blue),  6.7 wt.\% of 4H-SMO (green) and 4.2 wt.\% of Mn$_3$O$_4$ (red) phases in the SMO sample. Experimental diffraction pattern is shown in comparison with the calculated patterns. The difference between measured and calculated profiles is shown as a curve at the bottom of the diagram. Short vertical bars indicate the positions of diffraction maxima of 6H-SMO (upper blue), 4H-SMO (middle green) and Mn$_3$O$_4$ (lower red), respectively. Inset shows enlarged part of the pattern with the indexed diffraction lines of 6H- and 4H-modifications of SMO.}
\label{rys-1}
\end{figure}

X-ray powder diffraction measurements of the strontium manganite SMO showed multiphase composition of the sample. Besides the main 6H-SrMnO$_3$ (6H-SMO) phase the sample contains 4H-modifcation of SMO (4H-SMO) and minor amount of Mn$_3$O$_4$ hausmannite phase. It was confirmed by the precise high-resolution X-ray synchrotron powder diffraction studies carried out at beamline ID22 of ESRF. According to quantitative full profile Rietveld refinement (Fig. \ref{rys-1} ) the amount of parasitic 4H-SMO and Mn$_3$O$_4$ phases in the SMO sample is 6.7 and 4.2 wt. \%, respectively. In the refinement procedure, the unit cell dimensions, positional and displacement parameters of atoms in the 6H- and 4H-SMO structures were refined together with background and peak profile parameters, texture correction and correction of absorption and instrumental zero shift. For the minor Mn$_3$O$_4$ phase only the unit cell dimensions were refined. As the starting models for the refinement, the atomic positions in 6H- and 4H- polymorphs of SMO \cite{21,25} and in tetragonal Mn$_3$O$_4$ \cite{26} were used. Obtained structural parameters of both hexagonal modifications of SMO presented in the SMO sample and corresponding residuals are collected in Table \ref{tab-1}.

\begin{figure*}
\centering
\includegraphics[width=0.90\textwidth]{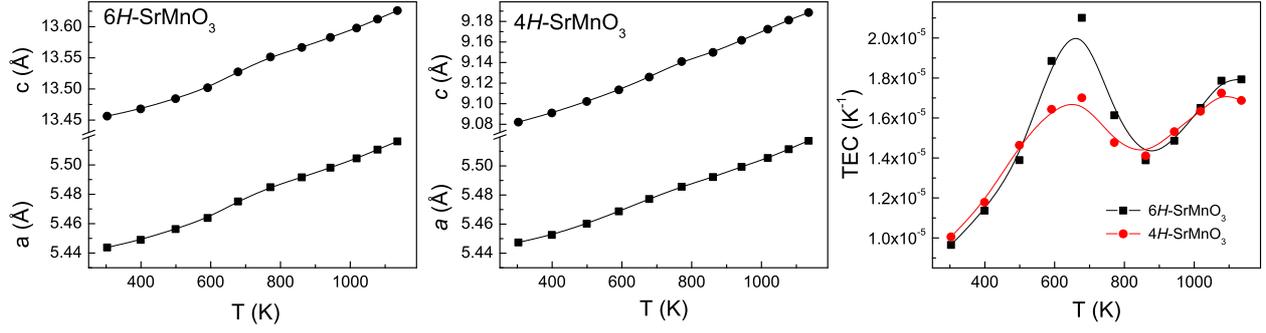}
\caption{Temperature dependencies of the crystal lattice parameters of  6H-SMO and 4H-SMO (left and middle panels) determined from XRD. Average  linear normalized temperature expansion coefficient (right panel) of these structures TEC were obtained by differentiation of experimental values of unit cell dimensions on the temperature. Solid lines are guide for eyes.}
\label{rys-3}
\end{figure*}

The main 6H-SMO phase detected in SMO sample contradicts with the phase diagram of the SrMnO$_{3-x}$ system reported in \cite{Neg-70,Neg-73}, according to which at this sintering temperature (1603 K) the 4H-SMO is a stable phase. However, structural instability of four-layer hexagonal SMO was reported \cite{Neg-70},
 which becomes increasingly anion deficient above 1308 K. Hexagonal 6H-modification of SMO is considered in the literature as a high-pressure phase, which is thermodynamically stable between 5.9(3) and 18.1(2) GPa at 0 K \cite{Niel-14}. 
According to the broad outline of the phases which should be observed at thermodynamic equilibrium in SMO system \cite{Niel-14}, the 6H-SMO is stable in a large PT region between 4H hexagonal and 3C cubic perovskite phases. However, it was emphasized that the exact location of the borders between different SMO polymorphs are still quite uncertain due to the intensive oxygen loses at the elevated temperatures, which strongly affects the thermodynamics of the compounds. It was concluded \cite{Niel-14} that further investigations are required in order to establish whether the stability field of 6H-SMO is expanded or contracted with increasing oxygen deficiency in SrMnO$_{3-x}$ system. Coexistence of the 6H- and 4H-polymorphs of SMO detected in our SMO sample proves very complex phase and structural behaviour in the SrMnO$_{3-x}$ system. It is evident, that peculiarities of the synthesis procedure (heating/cooling rates, temperature and duration of heat treatment and ambient atmosphere, quenching or slow cooling, etc.) as well as effect of some other factors like as uncontrolled impurities could influence on the phase composition of SMO specimens. 

\begin{table*}
\centering
\begin{tabular}{||c|c|c|c|c|c||}
\hline\hline
Lattice parameters [\AA] 	&	Atoms, sites	&	$x$/$a$	&	$y$/$b$	&	$z$/$c$	&	$B_{iso/eq}$ [\AA$^2$]	\\
\hline \hline
\multicolumn{6}{||c||}{6H-SMO (space group $P6_3/mmc$, $R_I$ = 0.022, $R_p$=0.203)}\\
\hline
$a$ = 5.4437(5)		&	Sr1, 2$b$	&	0			&	0	&	1/4			&	0.92(3)	\\
$c$ = 13.4564(9)	&	Sr2, 4$f$	&	1/3		&	2/3&	0.0885(3)	&	0.68(2)	\\
						&	Mn1, 2$a$	&	0			&	0	&	0				&	0.17(4)	\\
						&	Mn2, 4$f$	&	1/3		&	2/3&	0.8433(8)	&	0.71(3)	\\
						&	O1, 6$h$	&	0.5114(5)&	2$x$	&	1/4			&	2.37(12)	\\
						&	O2, 12$k$	&	0.8317(4)&	2$x$	&	0.0794(2)	&	0.88(5)	\\
\hline
\multicolumn{6}{||l||}{Texture axis		 parameter: [110]; 0.656(8)}\\
\hline 	\hline										
\multicolumn{6}{||c||}{4H-SMO (space group $P6_3/mmc$,  R$_I$ = 0.137, R$_p$=0.203)}\\
\hline											
$a$ = 5.4474(2)	&	Sr1, 2$a$		&	0			&	0		&	0			&	0.52(8)	\\
$c$ = 9.0829(4)	&	Sr2, 2$c$		&	1/3		&	2/3	&	1/4		&	0.36(10)	\\
					&	Mn, 4$f$		&	1/3		&	2/3	&	0.6117(6)&	0.33(10)	\\
					&	O1, 6$g$		&	1/2		&	0		&	0			&	0.3(3)	\\
					&	O2, 6$h$		&	-0.178(2)&	2$x$		&	1/4		&	0.4(3)	\\
%Texture axis;	& parameter:	& [110];		&			& 1.33(9)	&				\\
\hline
\multicolumn{6}{||l||}{ Texture axis;	 parameter:	 [110];					 1.33(9)}\\
\hline		\hline									
\multicolumn{6}{||c||}{15R-BMO (space group $R$-$3m$,  R$_I$ = 0.034, R$_p$=0.146)}\\
\hline											
$a$ = 5.6756(13)	&	Ba1, 3$a$	&	0	&	0	&	0	&	0.54(6)	\\
$c$ = 35.326(2)	&	Ba2, 6$c$	&	0	&	0	&	0.1337(2)	&	1.16(4)	\\
					&	Ba3, 6$c$	&	0	&	0	&	0.2656(4)	&	1.41(4)	\\
					&	Mn1, 3$b$	&	0	&	0	&	1/2	&	1.7(2)	\\
					&	Mn2, 6$c$	&	0	&	0	&	0.3622(1)	&	0.95(9)	\\
					&	Mn3, 6$c$	&	0	&	0	&	0.4312(2)	&	1.28(6)	\\
					&	O1, 9$e$	&	1/2	&	0	&	0	&	2.6(3)	\\
					&	O2, 18$h$	&	0.1891(8)	&	- $x$	&	0.0648(2)	&	0.8(2)	\\
					&	O3, 18$h$	&	0.4840(10)	&	- $x$	&	0.1333(3)	&	2.1(3)	\\
%Texture axis;	& parameter: &[001]	&; 1.413(8)& & \\
\hline
\multicolumn{6}{||l||}{ Texture axis;	 parameter: [001]; 1.413(8)}\\
\hline		\hline									
\multicolumn{6}{||c||}{4H-BSMO (space group $P6_3/mmc$, R$_I$ = 0.037, R$_p$=0.110)}											\\
\hline											
$a$ = 5.5532(2)	&	Ba/Sr*, 2$a$	&	0				&	0		&	0				&	0.88(2)	\\
$c$ = 9.1536(3)	&	Sr/Ba**, 2$c$	&	1/3			&	2/3	&	1/4			&	0.90(3)	\\
					&	Mn, 4$f$		&	1/3			&	2/3	&	0.6122(2)	&	0.67(5)	\\
					&	O1, 6$g$		&	1/2			&	0		&	0				&	0.52(10)	\\
					&	O2, 6$h$		&	-0.1828(5)	&	2$x$		&	1/4			&	0.73(11)	\\
\hline
\multicolumn{6}{||l||}{* 0.67(2)Ba + 0.33(2) Sr;  ** 0.66(2) Sr + 0.34(2) Ba }\\
\hline\hline
\end{tabular}
\caption{Crystallographic data obtained for 6H-SMO 4H-SMO, 15R-BMO and 4H-BSMO structures in studied samples at RT.}
\label{tab-1}
\end{table*}

\begin{figure*}
\centering
\includegraphics[width=0.9\textwidth]{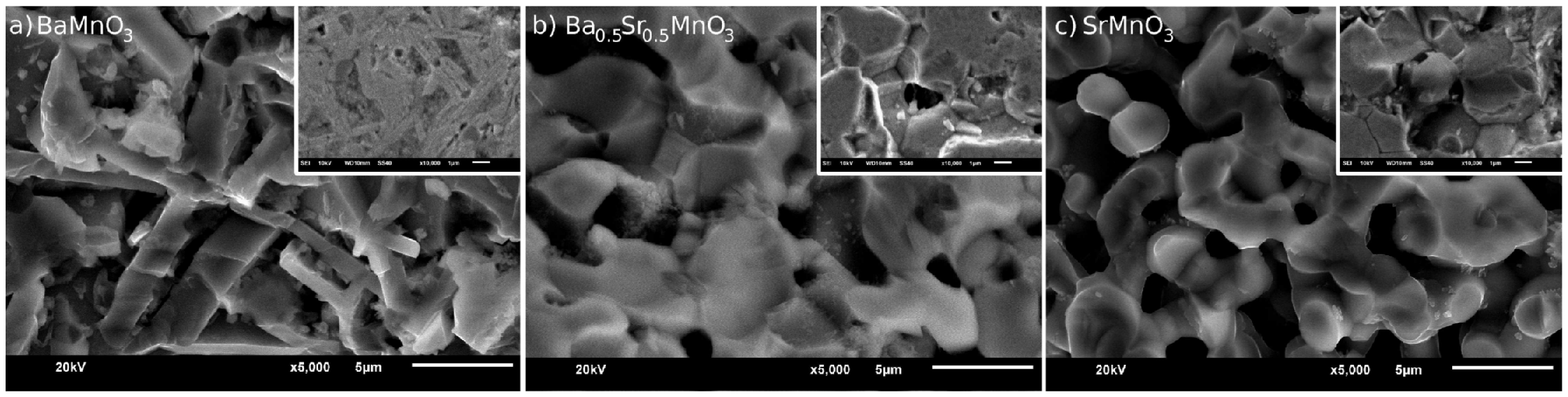}
\caption{The SEM microphotographs of the fractured surface, magnification 5000~$\times$: (a) BMO, (b) BSMO, (c) SMO. In the insets: The SEM microphotographs of the polished and etched samples surfaces,  magnification 10000~$\times$.}
\label{rys-4}
\end{figure*}

Analysis of XRD pattern of barium manganite (BMO) shown rhombohedral 15R-BMO modification as the main phase. Besides the main 15R-BMO phase in the amount of ~93 wt.\%, the sample contains $\sim$ 6 wt.\% of the 2H-BMO polymorph and minor amount of hausmannite Mn$_3$O$_4$ (less as 1 wt.\%) and unidentified phase(s). Corresponding percentages, as well as crystal structure parameters of the main 15R-BMO phase were obtained by simultaneous multiphase full profile Rietveld refinement. As starting models for the refinement, atomic positions in 15R-BMO \cite{27}, 2H- BMO \cite{28} and Mn$_3$O$_4$ \cite{26} were used. Obtained structural parameters for 15R-BMO structure (Table \ref{tab-1}) are in good agreement with the ones obtained for BMO single crystal in Ref. \cite{29}.

XRD pattern of the mixed strontium-barium manganite BSMO polycrystalline sample showed hexagonal 4H-BSMO polymorph as the main phase wit the estimated amount of  $\sim$ 94 wt.\%. In addition, several minor reflections of parasitic phases were detected, which can be assigned for 2H-BMO, Mn$_3$O$_4$ and extra unidentified phase(s). Crystal structure parameters of the main 4H-BSMO phase presented in Table  \ref{tab-1} were obtained by full profile Rietveld refinement by using atomic positions in BSMO \cite{30} as starting model for the refinement. The unit cell dimensions, positional and displacement parameters of atoms in BSMO structure were refined together with background and peak profile parameters, correction of absorption and instrumental zero shift as well as texture correction. 

The main structures detected in BMO and BSMO samples coincide well with the phase diagram of the Ba$_{1-y}$Sr$_y$MnO$_{3-x}$ system \cite{Neg-73}, according to which at those sintering temperatures the stable phases are the hexagonal 15R-BMO and 4H-BSMO.

\textit{In situ} high-resolution X-ray synchrotron powder diffraction studies of strontium manganite SMO in the temperature range of 298-1135 K presented in Fig. \ref{rys-3} showed no visible structural changes in both 6H- and 4H-polymorphs presented in the sample. Crystal structure parameters of both phases at the elevated temperatures were successfully refined in space group $P6_3/mmc$. However, an analysis of the temperature dependencies of the unit cell dimensions revealed remarkable anomalies in the lattice expansion of both phases. The anomalies are reflected in a kink of the lattice parameters at 600-800 K range that caused an anomalous increase of thermal expansion coefficients (TECs), with a clear maximum in a vicinity at 670 K (Fig. \ref{rys-3}).

Typically, the average linear TECs of Sr-containing manganites at 298 -1273 K are in the range of 11 -13$\times$10$^{-6}$~K$^{-1}$ \cite{31,32,33}. The abnormal increase of TEC observed in 6H-SMO  and 4H-SMO is related with the variation of the oxygen content and corresponding changes of the defect structure at the elevated temperatures. It has been recently reported for the related calcium manganite CaMnO$_{3-\delta}$ \cite{32} that an increase of the thermal expansion coefficient of CaMnO$_{3-\delta}$ near 1170 K was explained by the oxygen loss at the heating of the sample and formation of Mn$^{3+}$ species with the size larger than Mn$^{4+}$ cations.

Information on the change of defect structure of strontium manganite at the elevated temperatures was reported in \cite{34}. Increase in the anion deficiency of hexagonal 4H-SrMnO$_{3-\delta}$ in the temperature range of 620-670 K detected by TGA/DTA appears to be in good agreement with the conductivity variation showing the changing behaviour at the same temperature range \cite{34}. Our \textit{in situ} high-temperature structural investigations of SrMnO$_3$ favour the formation of anion vacancies at the elevated temperatures. However, to explain the nature of the observed anomalies in the lattice expansion of both hexagonal SMO polymorphs the additional studies of caloric, transport and magnetic properties are required. 

\subsection{SEM microphotographs}

The SEM microphotographs obtained at RT with magnification 5000~$\times$  for fractured surfaces are presented in Figure \ref{rys-4}. The grains have different dimensions within the range 2-5~$\mu m$. They are clearly visible in Figs.\ref{rys-4}(a-c)  and in the insets in the SEM microphotographs of the polished and etched samples surfaces with  magnification 10000~$\times$. They correspond to precipitation of the second phase determined in the X-ray measurements. Moreover, one can observe the  porosity, typical for the used method of sintering of our samples.

\subsection{Magnetic measurements}
\begin{figure*}[htb]  
\centering
\includegraphics[width=0.90\textwidth]{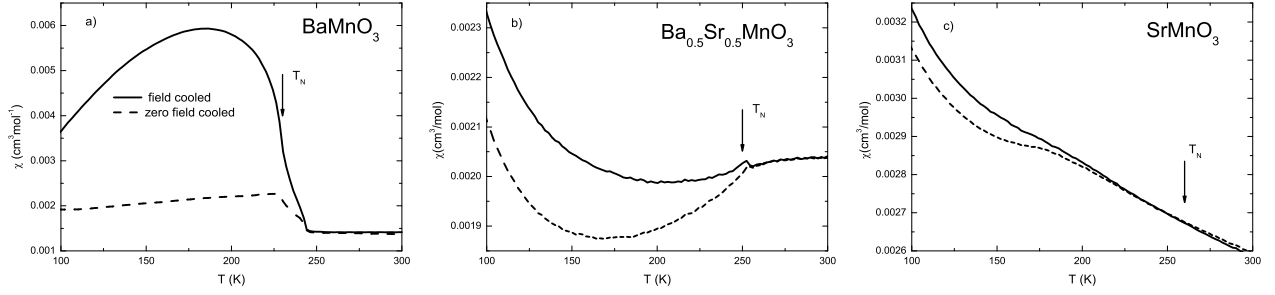}
\caption{Zero field cooling/field cooling susceptibilities at H = 50 Oe in the temperature range 100-300 K for $x = 1.0, 0.5$ and $0.0$ in the (a), (b), (c) plots, respectively.}
\label{rys-5}
\end{figure*}

	The system displayed a significant stoichiometry dependence of the magnetic properties, despite the lack of change in the number of the Mn magnetic centres within the unit cell. Fig. \ref{rys-5} presents the temperature dependence of the magnetic susceptibility $\chi$ of the samples measured in zero field
cooling (ZFC) and field cooling (FC) modes. Below 50 K an increase in susceptibility was observed (not shown) related to the parasitic Mn$_3$ O$_4$ phase, which was detected with XRD measurements. At room temperature all three samples are paramagnetic.Upon cooling, the anomalies in $\chi$, as well as the bifurcation of the ZFC and FC branches, point to the transitions to the antiferromagnetically ordered phase. The occurrence of paramagnetic to antiferromagnetic phase transitions was confirmed also by other techniques, including differential scanning calorimetry, the results of which are presented below. Based on Fig. 5 we determined the N$\acute{e}$el temperatures T$_N$ as follows: 230 K for BMO, 250 K for BSMO and 260 K for SMO ceramics. They are slightly different in comparison to the data for the
ceramics and crystals reported in literature \cite{12, 27, 29, 30}. The reason for these discrepancies is most probably the non uniphase character and structural modifications detected in all samples (see~\ref{sec3}.1) and \cite{35,36}. It was stated in \cite{27} that the ordering temperatures show a dependence on the particular
fraction of the cubic layers present in the staking sequence. One may expect that the multiphase nature brings about also the complex transition in BaMnO$_3$, where the main transition at 230 K is followed by the weaker one at 244 K (Fig. \ref{rys-5}a). There are two rather unusual features of the magnetic
susceptibility in the samples under study: the almost costant, non Curie-Weiss temperature dependence above T$_N$ and very small value at T$_N$. The similar behaviour was reported in \cite{27, 29, 30}. The authors of \cite{29} explain this fact by the strong magnetic anisotropy and confinement of the Mn spins in the basal plane.

\begin{figure}
\centering
\includegraphics[width=0.4\textwidth]{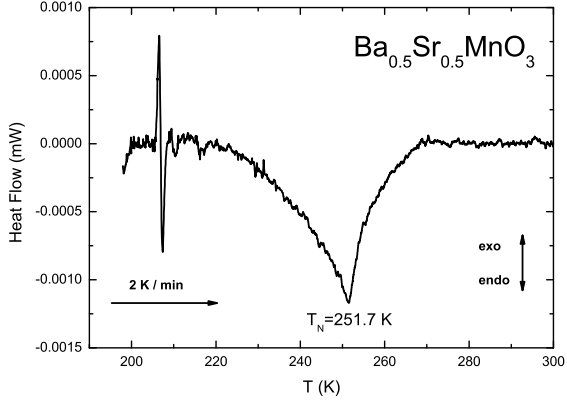}
\caption{DSC thermogram for BSMO obtained during heating.}
\label{rys-6}
\end{figure}

\subsection{DSC}
	The thermogram of BSMO is presented in Fig.~\ref{rys-6}.  One can observe the phase transition at 251.7 K which correlates very well with the N$\acute{e}$el  transition at 250 K for BSMO (Fig.~\ref{rys-5}b). For SMO and BMO the thermal behaviour was determined previously in  \cite{29,37}. The enthalpy and entropy changes of BSMO were estimated as 17.5 J/mol and 70~mJ/K mol, respectively. Around 205 K, the exo- and next endothermic peaks are visible; probably they are connected with structural changes at the investigated compound.

\subsection{FTIR test} 
\begin{figure}[htb] 
\centering
\includegraphics[width=0.4\textwidth]{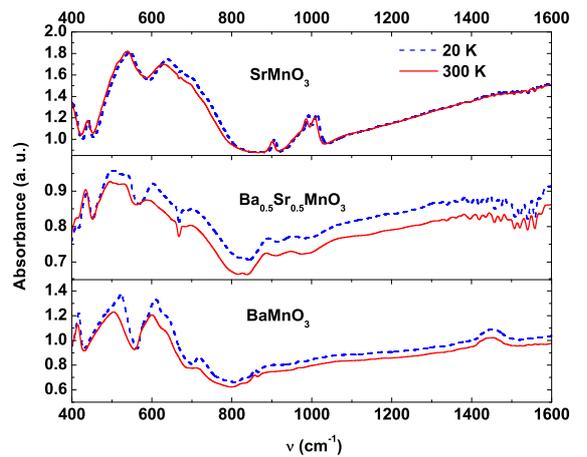}
\caption{The infrared absorption spectra detected at 20 K and 300~K for BMO, BSMO, and SMO.}
\label{rys-71}
\end{figure}

In Fig. \ref{rys-71} the infrared absorption spectra are presented (at 20 K and RT) in the frequency range from 400--1600~cm$^{-1}$ for SMO, BSMO and BMO. The observed vibrational bands for both temperatures are shown in Table \ref{tab-3}. 
These bands are related with stretching Mn-O and O-Mn-O deformational (in plane and out-of-plane) bands. For BSMO the temperature changes of the position of the  selected vibrational bands around 536, 829, 891 and 925~cm$^{-1}$ were observed. 
The band around 829~cm$^{-1}$ decreased to 825~cm$^{-1}$ when the temperature changed from 210 to 220 K. This behaviour is associated with the structural phase transition observed in the DSC diagram at about 205 K (Fig.~\ref{rys-6}), and next in the DMA results. The region of the changes of the vibrational band position around 598~cm$^{-1}$ in temperature range from 245 to 272 K overlaps with  the region of the transition between paramagnetic and antiferromagnetic phases 240-260~K. The frequency of this band decreased to 586~cm$^{-1}$.
We discovered the anomalies at the $T_N$ for BSMO in the different physical properties. Kamba \textit{et al.} observed multiphonon and multimagnon scattering in Raman spectra \cite{37}. But we cannnot discuss these ideas because of hexagonal structures and nonhomogeneity of our samples.  

\begin{table}
\centering
\begin{tabular}{||cc|cc|cc||}
\hline \hline 
\multicolumn{2}{||c|}{SMO} & \multicolumn{2}{c}{BSMO} & \multicolumn{2}{|c||}{BMO}\\
\scriptsize T=20K	&	\scriptsize T=296K	&	\scriptsize T=20K	&	\scriptsize T=296K	&	\scriptsize T=20K	&	\scriptsize  T=296K	\\
\scriptsize (cm$^{-1}$)	&	\scriptsize (cm$^{-1}$)	&	\scriptsize (cm$^{-1}$)	&	\scriptsize (cm$^{-1}$)	&	\scriptsize(cm$^{-1}$)	&	\scriptsize (cm$^{-1}$)	\\
\hline
421	&	421	&	440	&	436	&	440	&	440	\\
521	&	505	&	498	&	498	&	544	&	540	\\
610	&	602	&	602	&	586	&	636	&	625	\\
718	&	0	&	702	&	698	&	907	&	903	\\
1450	&	1454	&	891	&	887	&	991	&	986	\\
0	&	1701	&	953	&	949	&	1013	&	1009	\\
\hline \hline 
\end{tabular}
\caption{Infrared absorption spectra measured at 20 K and RT in frequency range from 400--1600~cm$^{-1}$ for SMO, BSMO, and BMO. Peaks position are noted.}
\label{tab-3}
\end{table}

\subsection{Dielectric measurements} 

\begin{figure}
\centering
\includegraphics[width=0.4\textwidth]{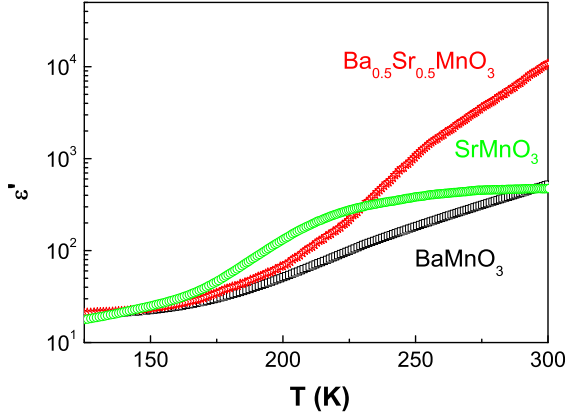}
\caption{The real part $\varepsilon '$ of the dielectric permittivity as a function of the temperature measured at $f$=100 kHz for BMO, BSMO and SMO.}
\label{rys-8}
\end{figure}

The real part of the dielectric permittivity temperature dependence $\varepsilon '$($T$) measured at 100 kHz for BMO, BSMO and SMO is presented in  Fig.~\ref{rys-8}. All investigated samples show an increase in the dielectric permittivity $\varepsilon '$ above the temperature of 150 K.

The sample of SMO shows a step in dielectric permittivity, in 170-220 K range. This effect correlates with the peak in the DSC measurements for this compound ( compare Figs.~\ref{rys-6} and ~\ref{rys-8}). A weak step in  $\varepsilon '$($T$), hardly discerned, occurs for BSMO and BMO samples.

\begin{figure}
\centering
\includegraphics[width=0.3\textwidth]{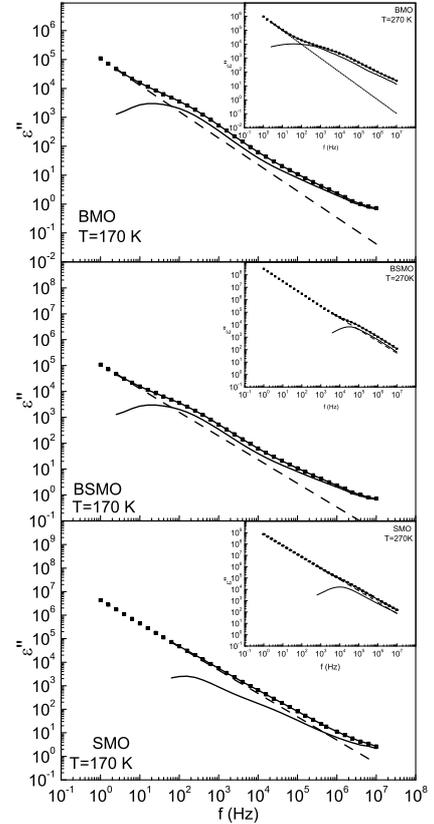}
\caption{The imaginary part  $\varepsilon "$ of the dielectric permittivity as a function of the frequency $f$  measured at 170 and 270 K for BMO, BSMO and SMO; continuous lines over the experimental points represent fits to the Cole-Cole equation; the dashed line is related to the conductivity term $(\sigma_0/ \varepsilon_0 \omega)^N$; two other continuous lines represent the contribution of the grain boundary (270 K) and grain interior (170 K).}
\label{rys-9}
\end{figure}

Analysis of dielectric losses $\varepsilon "$ spectra, presented in Fig. \ref{rys-9}, provides more details about the dielectric response. The temperatures 170 and 270~K were chosen as the representatives for the low temperature ($\sim$ 140--240 K) and the high temperature (250--300~K) relaxation processes caused by mechanism of polarons and/or  oxygen vacancies conductivity. The experimental results of the dielectric spectrum measurement were fitted with the Cole--Cole equation \cite{40} superposed to the conductivity term \cite{41} using the WinFit V 3.2 computer program:

\begin{equation}
\epsilon(\omega) = \varepsilon ' - i \varepsilon "=
-i \left( 
\frac{\sigma_0}{\varepsilon_0 \omega}\right)^N +
\sum_{k=1}^2\left[  
\frac{\Delta \varepsilon_k}{1+ (i \omega \tau_k)^\alpha}
\right] + \varepsilon_{\infty k}
\label{eq1}
\end{equation}

where: $k$= 1, 2 - number of relaxation processes, $\Delta$  - difference between $\varepsilon '$ values  at low frequency and high frequency limit,   $\varepsilon_{\infty}-$\ value of $\varepsilon '$  at high frequency limit, $\alpha$ - degree of the distribution of relaxation time $\tau$,  $\sigma_0$ - specific dc conductivity, $N$ - exponent of the frequency dependence of $\varepsilon "$. In the Fig. 10 the black square points represent the experimental data which are in good agreement with the best fits to the Cole - Cole equation, represented by the continuous lines over the points. The dashed line is related to the dc conductivity term  $(\sigma_0/ \varepsilon_0 \omega)^N$  in the Eq.~\ref{eq1} ($k$= 1). The other continuous lines refer to the contribution of relaxation processes (k=2).

\begin{figure}[htb]
\centering
\includegraphics[width=0.40\textwidth]{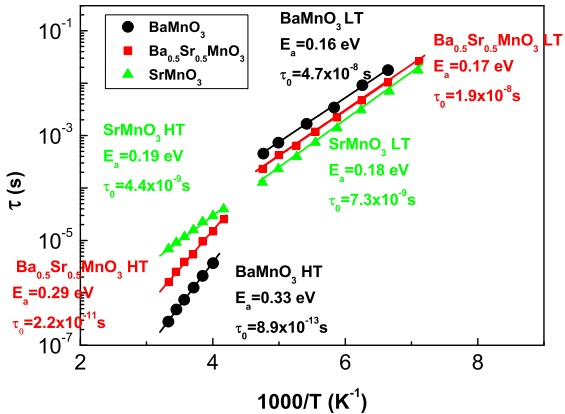}
\caption{The relaxation times of two visible processes versus reciprocal temperature obtained for all investigated samples. The activation energy $E_a$ values were calculated from Eq. \ref{eq2} for the low (LT) and high temperature (HT) regions.}
\label{rys-10}
\end{figure}

The activation energy $E_a$ characteristic for relaxation processes can be determined from the Arrhenius law:

\begin{equation}
\tau = \tau_0 \exp (E_a/k_B T)
\label{eq2}
\end{equation}

where $\tau$ denote the relaxation time,  $\tau_0$ as the preexponential factor is characteristic relaxation time, and $k_B$ is the Boltzmann constant. Fig.~\ref{rys-10} shows the relaxation times of processes versus reciprocal temperature obtained for all investigated samples. The shortest values of $\tau_0$ indicate possibility of polarons involved in the relaxation process in high temperature regions. The activation energy $E_a$ values calculated from Eq.\ref{eq2} are also given in Fig.~\ref{rys-10}. It is worth to note that the $E_a$ values, obtained from the dielectric measurements, are similar to these ones determined in the impedance spectra study \cite{16}. Therefore electric nonhomogeneity can be deduced.

\begin{figure}[htb] 
\centering
\includegraphics[width=0.40\textwidth]{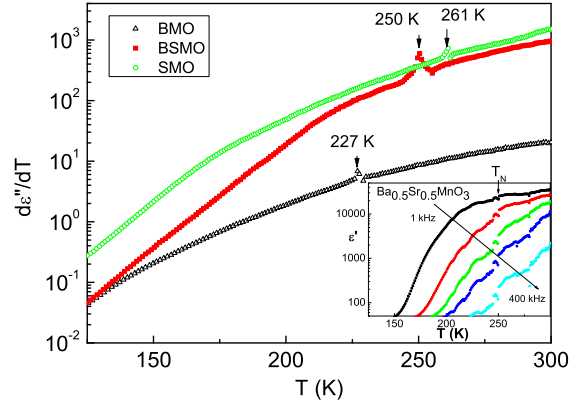}
\caption{Plots of the derivative $d\varepsilon "/dT$ versus temperature for BMO, SMO and BSMO. In the inset the thermal dependence of the dielectric permittivity for chosen frequencies for BSMO ceramic - there is visible the anomaly at 250~K connected with N$\acute{e}$el phase transition for frequencies 1, 10, 40 100, 400~kHz.}
\label{rys-11}
\end{figure}

In Fig.~\ref{rys-11}, the derivative $d\varepsilon "/dT$ versus temperature plots show small peaks at the temperatures 227, 261 and 250 K for BMO, SMO and BSMO, respectively. The positions of the peaks are in a good agreement with the $T_N$ temperatures assigned in the magnetic experiment Fig. ~\ref{rys-5}. 
Therefore, we can expect weak connection between magnetic and dielectric properties that will be the subject of our future studies; however, there should be made an effort to settle the discussion if the effects are caused by parasitic phases or not, especially in BSMO.

\subsection{DMA test} 
\begin{figure}[htb]
\centering
\includegraphics[width=0.3\textwidth]{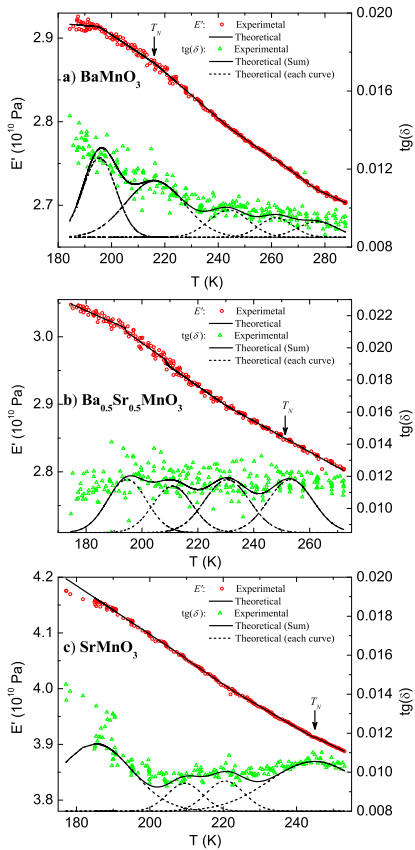}
\caption{ Results of DMA. Experimental (open circles) storage modulus $E'$ and tangent of losses versus temperature $T$ for (a) BMO, (b) BSMO, (c) SMO with fitted linear dependences of $E'$($T$) (solid lines), non-normalised Gaussian functions (dashed curves) and sums of Gaussian functions (solid curve). N$\acute{e}$el tempeartures $T_N$ estimated by other techniques are dentoes by arrows.}
\label{rys-12}
\end{figure}
The storage modulus $E'$ for empty steel pockets in the broad temperature range is a straight line decreasing with increasing temperature. There is no anomaly observed between liquid-nitrogen up to RT temperature.
Figures ~\ref{rys-12}(a-c) show the experimental and theoretical values of the storage modulus $E'$ and the tangent of mechanical losses $tan(\delta)$ versus temperature $T$ for BMO, BSMO and SMO, respectively. In the measurements was applied small dynamical shear which caused small deformation as a response of the system for example see \cite{19}. Therefore, due to the Hook's law, the linear dependence between shear and deformation is expected there. For that reason to the experimental data of $E'$ the straight lines have been fitted with various values of the slopes. As long as there is no possibility to fit a single line for the experimental data shown in Figs. \ref{rys-12} a-c, the stright lines have been fitted in the preliminarily selected ranges of temperatures (not broader than 7-10 K). Then these regions have been extended (by including one-by-one of experimental points) and the chi-squared test has been calculated in order to get minimal numbers of regions with variuos slopes of $E'$($T$). Moreover, only these regions have been left for which the slopes of neighbouring regions differ more than 10\%. The loss tangent as a function of temperature should be given by Gaussian or Lorentzian functions with maximum at the phase transition or glass transition points \cite{19}. Therefore, to the experimentally obtained $tan(\delta)$ the non-normalized Gaussian functions were fitted with various values of amplitudes and distribution parameters. The temperature of the maxima positions, fitted with the use of non-normalized Gaussian functions, are the same (or differ no more than 3K) as temperatures of intersections of fitted lines to storage modulus. 
Figure \ref{rys-12}a shows the temperature dependences of the storage modules  $E'$ and mechanical losses $tan(\delta)$ for BMO sample. The temperatures of the highest changes of slopes of $E'$, and the temperature locations of maxima of the highest losses - $tan(\delta)$ - are equal to about 196 K and 216~K. The former one can be equated with some structural phase transition and, additionally, it is close to the temperature of the maximum in the zero-field cooled magnetic susceptibility (compare Fig.~\ref{rys-5}a). Additionally, it is a temperature of a start of the pre-transiotional effect seen as the beginning of the lambda peak detected by calorimetry measurements \cite{29}. The latter temperature is close to the N$\acute{e}$el temperature found by magnetic and DSC measurements and reported in \cite{30}. Several additional regions with different values of the slopes of $E'$ and characterized by low-amplitude Gaussian peaks of $tan(\delta)$ are visible at temperatures about 244 K, 262 K, and 277 K, the first one agreeing with the phase transition visible in the zfc/fc curves (Fig.~\ref{rys-5}a). 
Figure \ref{rys-12}b presents the dependences of $E'$ and $tan(\delta)$ on temperature $T$ for the BSMO sample. Two anomalies at about 195 K and 209 K accompanied by step-like behaviour of $E'$ were observed. They denote structural phase transitions. Obtained temperatures of these transitions are close to adequate ones seen on: the DSC diagram at 205 K (see Fig.~\ref{rys-6}), the location of vibrational band around 829 cm$^{-1}$ obtained from FT-IR measurements at 213 K, and slope change in the zero-field cooled susceptibility - between 195 and 212 K (see Fig.~\ref{rys-5}b). The next temperatures of slope changes of $E'$ and location of the maximum of $tan(\delta)$ are equal to about 231 K and 250~K. The former one is the same as on-set temperature of DSC peak (see Fig. 6), similar as for BMO sample. The latter one is equal to $T_N$ obtained from DSC diagram, magnetic measurements and FT-IR and dielectric spectroscopies - see Figs. 5, 6, and 10, respectively. The amplitudes for fitted non-normalized Gaussian functions are almost the same. Pre-transitional effect means start of a structural changes, reinforcement or weakening of interactions between atoms, vacancies creation or migration, etc., that can lead to change in mechanical properties. Moreover, near $T_N$ the lattice is extraordinary 'soft' and susceptible to even slight displacement of oxygen, manganite or barium atoms \cite{29}. The temperature dependences of the storage modules $E'$ and mechanical losses $tan(\delta)$ are shown in Fig. 11c for SMO sample. The four regions of various slopes were detected. Two of them are characterized by significant changes of the slopes of $E'$ and, at the same time, high amplitudes of non-normalized Gaussian functions - for temperatures equaling about 185 K and 245 K. The first one is a certain structural phase transition, seen in the temperature dependence of the zero-field cooled magnetic susceptibility as the inflection point, about 175 K (see Fig. \ref{rys-5}c). The higher one is close to the N$\acute{e}$el temperature $T_N$ obtained from magnetic (260~K, Fig. \ref{rys-5}c) and dielectric (261.1 K, Fig. \ref{rys-11}) measurements. This temperature falls within the reported N$\acute{e}$el temperatures 230 K for SMO crystal \cite{13, 19} and 260 K for SMO polycrystal~\cite{35}. As long as near  $T_N$ the lattice is susceptible to slight displacements of any atoms building the structures of BMO, SMO and BSMO (for example see \cite{29}), the presence of other polymorphs, parasitic phases, functional groups, etc. (for example see section 3.1), the changes of behaviour of $E'$ on $T$ can be expected. Together with increasing temperature the interactions between additional minor 'inclusions' and 'host' lead to subtle chnages of $E'$($T$). It is worth to point it out, that determined tempeartures of changes in slopes of $E'$($T$) correspond well with anomalies and/or changes in behaviour found by other (complementary) techniques reported hearin and in \cite{24,37}.

\section{\textit{Ab initio} simulations results } 
\label{sec4}

\begin{figure}
\centering
\includegraphics[width=0.25\textwidth]{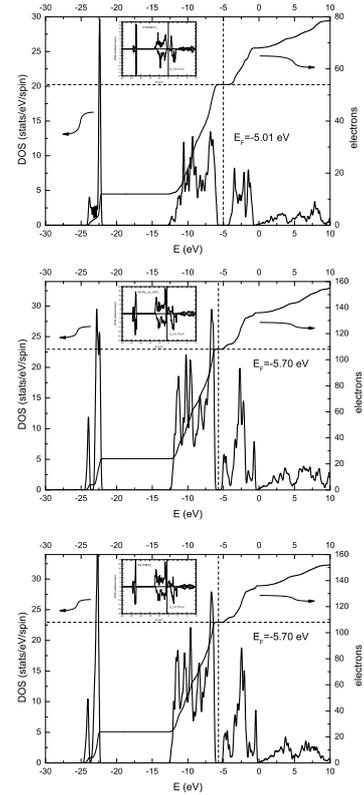}
\caption{Total density of states with signed integral of DOS by a grey line  (it visualizes a method of calculation of $E_F$). In the insets: Partial density of states for 2H-BMO, 4H-BSMO and 4H-SMO.}
\label{rys-14}
\end{figure}

Pseudo-potentials for Ba, Sr, Mn, O atoms were taken from Siesta GGA Pseudopotential Database. The following atomic configurations and cut-off radii: 
Ba: 6s$^2$ r= 4.37~$a_B$,  6p$^0$  r= 5.08~$a_B$, 5d$^0$ r= 4.05~$a_B$, 4f$^0$  r= 3.28~$a_B$, 
Sr: 5s$^2$  r=3.58~$a_B$, 5p$^0$ r= 4.11~$a_B$, 4d$^0$ r= 3.20~$a_B$, 4f$^0$ r= 3.58~$a_B$, 
Mn: 4s$^2$ r= 2.51~$a_B$, 4p$^0$ r= 2.57~$a_B$, 3d$^5$  r= 2.35~$a_B$,  4f$^0$ r= 2.35~$a_B$, 
O:  2s$^2$  r= 1.47~$a_B$, 2p$^4$ r= 1.47~$a_B$, 3d$^0$ r= 1.47~$a_B$, 4f$^0$ r= 1.47~$a_B$,
 and based on initial tests \cite{20}, a cut energy as 650 Ry, and the 8~$\times$8~$\times$8 Monkhorst-Pack grid for the Brillouin zone integration were used.  Starting from X-ray data at RT we fixed atomic positions and crystal lattice parameters we simulated the density of states (DOS) and a band gap energies ($E_{gap}$) for the stoichiometric non defected monocrystals of BMO, BSMO and SMO. Next, we carried out the calculations for the real SMO crystal with the hexagonal structure using lattice parameters which were found from temperature depending X-ray measurements. 
The elementary cell were simulated for four possible different configurations of atoms Ba and Sr positions in hexagonal 4H unit cell. There almost the same results were obtained. Therefore, as a representative configuration we chose this one with the lowest steady state energy. All formula units simulations were performed with spin polarization. For SMO we simulated 4H and 6H phases -Fig.\ref{rys-14}a, 4H for BSMO - Fig.\ref{rys-14}b, and 2H for BMO - Fig.\ref{rys-14}c. The structure 2H - BMO is an approximation of the 15R real structure analyzed in \cite{28}. 

In he relaxation processes of the unit cells for  ideal structures of 2H-BMO, 4H-BSMO, 4H-SMO, and 6H-SMO  there were obtained lattice parameters as follows: 2H-BMO $a$=$b$=5.746 (\AA), $c$=4.863 (\AA), $\alpha$=$\beta$=90.000 ($^o$), $\gamma$=119.987 ($^o$); 4H-BSMO  $a$=$b$=5.572 (\AA), $c$=9.130 (\AA), $\alpha$=$\beta$=89.999 ($^o$), $\gamma$=119.997 ($^o$); 4H-SMO $a$=$b$=5.427 (\AA), $c$=9.060 (\AA), $\alpha$=$\beta$=90.000 ($^o$), $\gamma$=119.992 ($^o$); 6H-SMO $a$=$b$=5.422 (\AA), $c$=13.374 (\AA), $\alpha$=$\beta$=90.000 ($^o$), $\gamma$=119.997 ($^o$).

	Based on our simulations within SIESTA 3.2 program we made DOS diagrams for 2H-BMO, 4H-BSMO, and 4H-SMO ideal stoichiometric crystals at 0 K. They are presented in Fig. \ref{rys-14}. One can see that the level of Fermi energy $E_F$ in these materials is about 5 eV and values of band energy gap  are 
1.49, 0.66, and 0.79 eV, respectively. $E_F$ was calculated by filling the complete DOS by electrons in the same amounts as this one in pseuodopotentials (eg. for 2H-BMO we have Ba: 6s$^2$, Mn: 4s$^2$, 3d$^5$, O: 2s$^2$, 2p$^4$ it means the sum 2+7+3x6=27e/f.u., for two formula units we get 54 electrons (see Fig. \ref{rys-14})). In analogical manner there were obtained $E_F$ for 4H-BSMO and 4H-SMO. $E_F$ is placed in the half of the gap energy gap. Typical problems in DFT theory are existence and a value of an energy gap and its deviation compared  with experimental results, which were discussed by many authors, ${e.g.}$ in \cite{42, Kor-14}. Observed differences in our measurements (see Fig. \ref{rys-10}) and in theoretical approximations are caused by heterogeneity of the samples such as presence of the parasitic phases and defects (especially oxygen vacancies).

	The contribution of the exchange correlation energy to the total energy is about 32~\%. Hence, magnetic properties in these systems are comparable. This fact is in good agreement with the experimental data - see Fig. \ref{rys-11} where the observed differences between systems are on the same level. To the comparison, the analogical parameters were previously simulated by the same method for BiMnO$_3$ in \cite{42}. There were achieved the values about 23 \%. This fact allows us to search a  coupling between magnetic and electric properties.

\section{Summary}
\label{sec5}  

X-ray powder diffraction measurements of the BMO, BSMO  and SMO ceramics revealed the structures as follows: 15R-BMO, 4H-BSMO  and 6H-SMO. The SMO shows the anomaly in the lattice expansion at 600-800 K with a maximum of thermal expansion coefficients at 670 K,  which is probably caused by an increase of anion oxygen deficiency and needs more studies. DMA measurements of storage modulus (Young modulus) of powder materials have shown that powder ceramic materials held in a steel pockets respond in a similar way as adequate bulk ones. As it has been shown, the powder DMA measurements clearly exhibit the changes in behaviour of mechanical properties of ceramics. Moreover, the indicated temperatures of changes of mechanical properties (storage modulus) are in good agreement with other data (DSC, dielectric spectroscopy and magnetic). The dependences of storage modulus $E'$ on temperature can be a continuous function as for BMO sample and for some transitions in BSMO and SMO samples. There is a connection between FTIR and DMA measurements for BSMO at 250~K - there are visible changes for phonon at 598~cm$^{-1}$ which shifted to 586~cm$^{-1}$ in the phase transition between paramagnetic and antiferromagnetic phases. The basic physical measurements such as magnetic studies, DSC, dielectric spectroscopy, FTIR and DMA for BSMO showed coupling between electric, magnetic and mechanical properties. Probably in this material, we have been  observed the elasto-electro-magnon but for confirmation next more advanced measurements are needed. Magnetic investigations confirmed paramagnetic-antiferromagnetic nature (Ba,Sr)MnO$_3$ system. There were found  the N$\acute{e}$el phase transition temperatures for studied ceramics at 230 K for BMO, 250 K for BSMO, and 260~K for SMO ceramics. Results of ${ab~initio}$ simulations  showed that the contribution of the exchange correlation energy to the total energy is about 32 \%. 

\section*{Acknowledgements}
The authors acknowledge the CPU time allocation at Academic Computer Centre CYFRONET AGH in Cracow. This work was supported in part by PL-Grid Infrastructure and the European Regional Development Fund under the Infrastructure and Environment Programme [grant number  UDA-POIS.13.01-023/09-00].  The  research  was  partially carried out with the equipment purchased thanks to the financial support of the European Regional Development Fund in the framework of the Polish Innovation Economy Operational Program (contract no. POIG.02.01.00-12-023/08). L. Vasylechko acknowledges partial support of the Ukrainian Ministry of Education and Sciences under the Projects "RZE", "KMON", and ICDD Grant-in-Aid program. The authors acknowledge Marek Kubik and Tomasz Prociak (from the Stanmark company, Cracow, Poland) for making available DMA measurements. The authors thank A. Fitch and Yu. Prots for kindly assistance with high-resolution synchrotron powder diffraction measurements at ID22 of ESRF during the beamtime allocated to the Experiment MA-2320 and prof. Stanislaw Baran (Institute of Physics, Jagiellonian University, Cracow, Poland) for a fruitful discussion.

%\section*{References}

\end{document}